\documentclass[12pt,twoside,usenatbib]{article}

\usepackage{xrb2007}
\usepackage{graphicx}
\newcommand{\LX}{\ensuremath{L_{\mathrm{X}}}}

\newcommand{\NH}{\ensuremath{N_{\mathrm{H}}}}

\newcommand{\erg}{\ensuremath{\mbox{erg}}}

\newcommand{\cm}{\ensuremath{\mbox{cm}}}

\newcommand{\cmsq}{\ensuremath{\cm^2}}

\newcommand{\nm}{\ensuremath{\mbox{\nm}}}

\newcommand{\ps}{\ensuremath{\s^{-1}}}

\newcommand{\s}{\ensuremath{\mbox{s}}}

\newcommand{\ergps}{\ensuremath{\erg~\ps}}

\newcommand{\CHANDRA}{\emph{Chandra}}

\begin{document}

\session{ULXs}

\shortauthor{Brassington}
\shorttitle{The LMXB Population of NGC 3379}

\title{LMXBs in the Normal Elliptical Galaxy NGC 3379}
\author{N. J. Brassington}
\affil{Harvard-Smithsonian Center for Astrophysics, 60 Garden Street, Cambridge, MA 02138}

\begin{abstract}
Presented here are the highlights from the deep \CHANDRA\ observation of the elliptical galaxy NGC 3379. From the multi-epoch observation of this galaxy, 132 discrete X-ray sources have been detected within the region overlapped by all observations, 98 of which lie within the D$_{25}$ ellipse of the galaxy. Of these 132 sources, 71 exhibit long-term
variability, indicating that they are accreting compact objects. 11 of
these sources have been identified as transient candidates, with a further 7
possible transients.
In addition to this, from the joint {\em Hubble/Chandra} field of view, nine globular clusters (GCs) and 53 field low mass X-ray binaries (LMXBs) have been detected in the galaxy. Comparisons of these two populations reveals that, at higher luminosities the field LMXBs and GC-LMXBs are similar. However, a significant lack of GC-LMXBs has been found at lower luminosities, indicating that not all LMXBs can form in GCs.

\end{abstract}

\vspace{-1.2cm}

\section{Introduction}
Low-mass X-ray binaries are the only direct fossil evidence of
the formation and evolution of binary stars in the old stellar
populations of early-type galaxies. First discovered in the Milky Way
(see \citet{Giacconi_74}), 
the origin and evolution of Galactic LMXBs has been the subject
of much discussion, centered on two main evolution paths (see
\citet{Grindlay_84}; review by \citet{Verbunt_95}): the evolution of
primordial binary systems in the stellar field, or formation and
evolution in Globular Cluster. 

With the advent of {\em Chandra}, many LMXB
populations have been discovered in early-type galaxies (see review
\citet{Fabbiano_06a}), and the same evolutionary themes (field or GC
formation and evolution) have again surfaced. However, most
observations have consisted of fairly shallow individual
snapshots for each observed galaxy, with limiting luminosity
of a few $10^{37}$~erg~s$^{-1}$. It is important to
study these old populations down to typical luminosities of the
Galaxy and M31. For this reason, alongside the need to identify the
variability of LMXB populations, we proposed (and were awarded) a very
large program of monitoring observations of nearby elliptical galaxies
with {\em Chandra} ACIS-S3. 

NGC 3379, in the nearby poor group Leo (D=10.6~Mpc)
was chosen for this study because it is a relatively isolated
unperturbed `typical' elliptical galaxy, with an old stellar
population and a poor globular cluster system. These characteristics
make NGC 3379 ideal for exploring the evolution of LMXB from primordial
field binaries.

\vspace{-0.4cm}

\section{The Catalog}

NGC 3379 was observed by \CHANDRA\ in five separate observations, carried
out over a six year baseline, with the first of these, a 30-ks pointing, being performed in February 2001. This observation has been followed by four
deeper pointings, all carried out between January
2006 and January 2007, resulting in a total exposure time of 337-ks.

From this co-added observation, 132 individual X-ray sources were detected within
the region overlapped by all observations, with 98 of these sources
lying within the D$_{25}$ ellipse of the galaxy. From these source
detections, source counts were extracted and fluxes, luminosities and
colors were calculated. Detailed analysis of this work, along with the full
catalog, is presented in \citep{Brassington_07a}, where flux and
spectral variability are also investigated.
In addition to the X-ray observation detailed in this catalog paper, optical data
are also presented, and X-ray/optical correlations are listed.

\begin{figure}

\begin{centering}
  \begin{minipage}{0.32\linewidth}
\includegraphics[width=\linewidth]{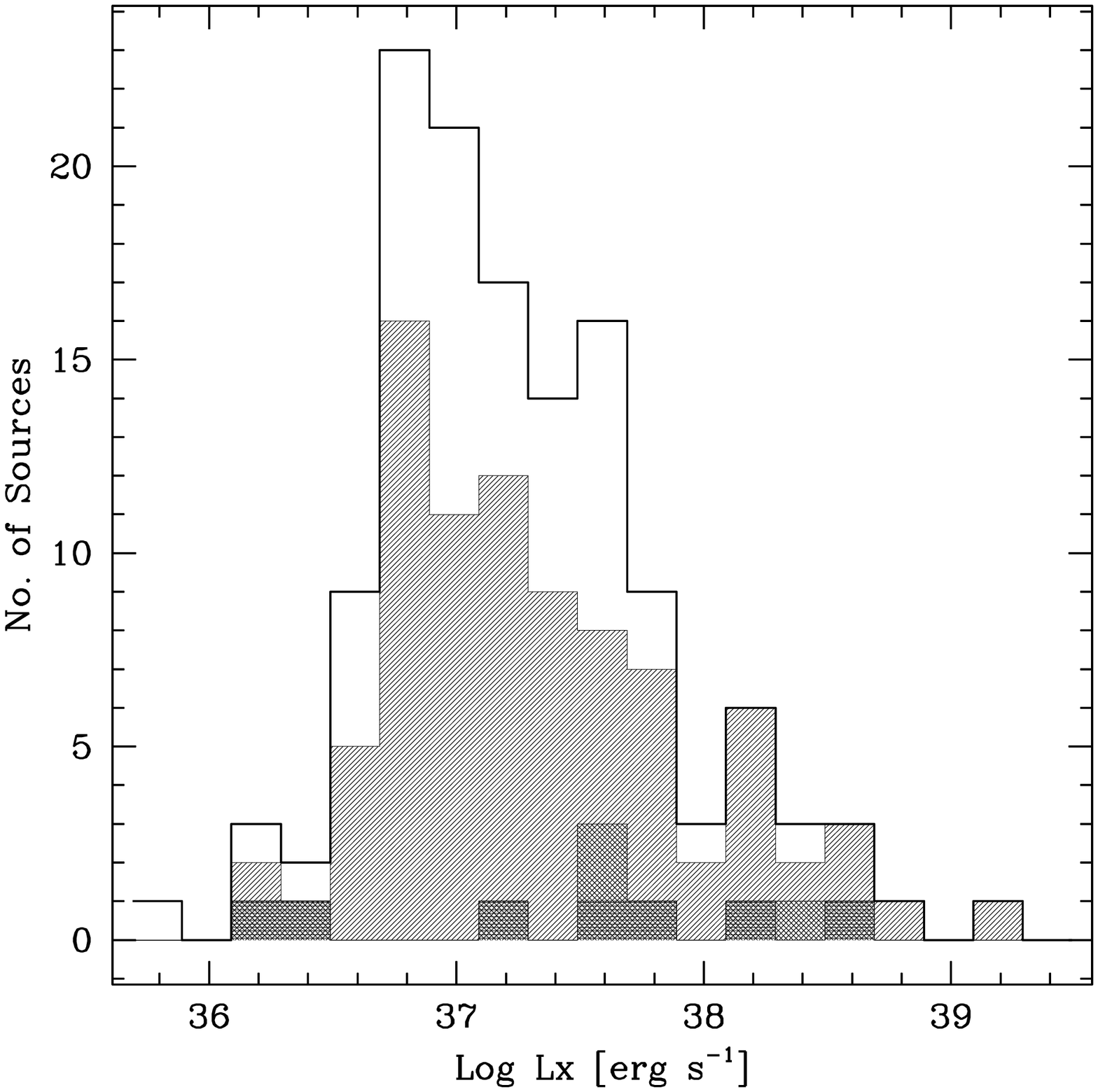}
\caption{}\label{fig:lxhist}
\end{minipage}
  \begin{minipage}{0.32\linewidth}
\includegraphics[width=\linewidth]{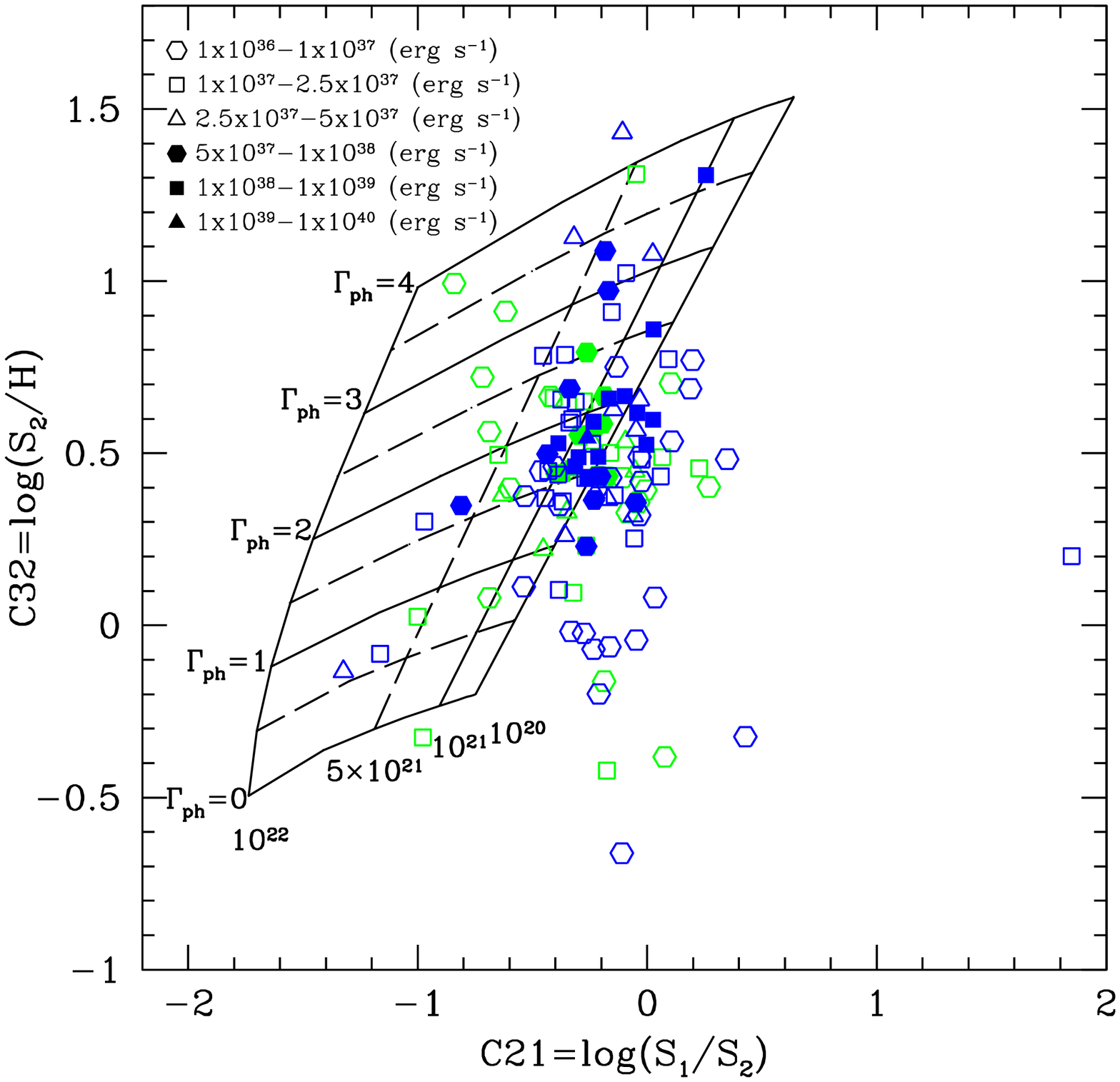}
\caption{}\label{fig:ccpop}
\end{minipage}
  \begin{minipage}{0.32\linewidth}
\includegraphics[width=\linewidth]{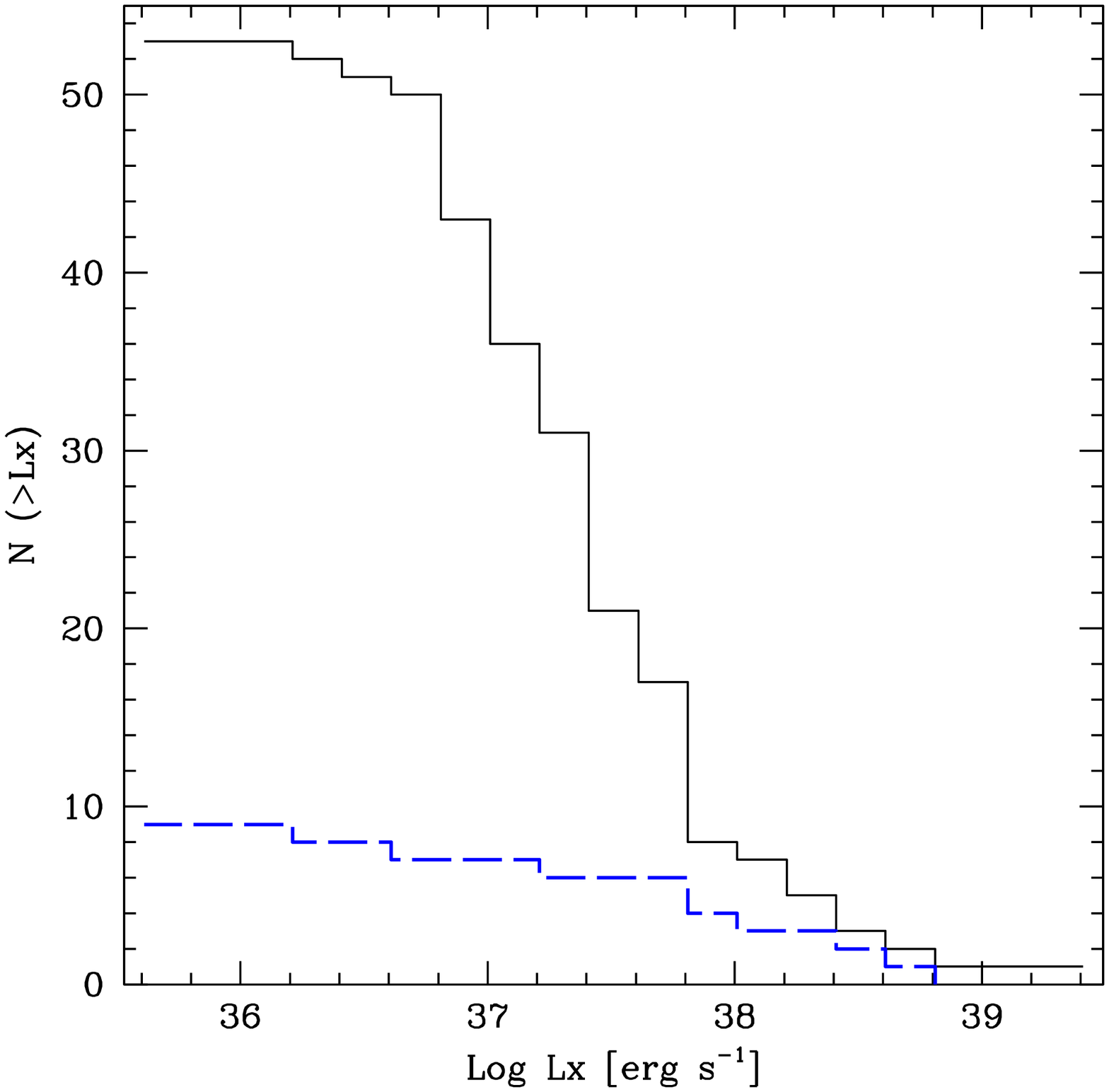}
\caption{}\label{fig:lf}
\end{minipage}
\end{centering}
\footnotesize{{\bf Figure 1:} \LX\ distribution of the 132 X-ray sources detected within the
overlapping region covered by all five \CHANDRA\ pointings. The unshaded histogram
indicates sources that have been determined to be non-variable
sources, with no GC counterpart. The lightly shaded region shows
variable sources that have no GC
counterpart. The darker histogram indicates non-varying sources
associated with a GC and the darkest histogram shows varying sources
that have a confirmed GC counterpart. {\bf Figure 2}: Color-color diagram of the X-ray point sources detected
in the co-added observation. Blue points indicate variable sources, and
green denotes non-variable sources. {\bf Figure 3:} Observed cumulative XLFs of field (solid black line) and GC
(dashed blue line) LMXBs; both distributions are from the joint {\em
Hubble/Chandra} field.  }
\end{figure}

\vspace{-0.4cm}

\section{Characterizing the LMXB Population}
The X-ray luminosity of the sources detected within NGC 3379 ranges from 6$\times~10^{35}$~\ergps\ up to $2\times10^{39}$\ergps,
where the brightest source is the only ULX detected within this
system. The properties of this source from the first two observations
have been reported in \citet{Fabbiano_06b}, and the analysis of the
full data set of the ULX will be
presented a forthcoming paper. The \LX\ distribution of 
all of the detected X-ray sources within NGC 3379 is shown in Figure
\ref{fig:lxhist}, where GC correlations are also shown. From this figure it can be seen that the majority of
sources detected from this observation lie in the luminosity range of
$5\times10^{36}$\ergps$-5\times10^{37}$\ergps, with a mode luminosity
of $\sim6\times10^{36}$\ergps.
Figure \ref{fig:ccpop} presents the LMXB population color-color diagram, based on the
photometry of the co-added observation. Here,
sources are divided into luminosity bins, with symbols of each
bin indicated by the labeling in the panel. Also in this figure source
variability is shown, where blue points indicate variable objects and green denotes
non-variable ones. Additionally, a grid has 
been overlaid to indicate the predicted 
locations of the sources at redshift $z$=0 for different spectra,
described by a power law with various photon indices
(0$\le\Gamma_{ph}\le4$, from top to bottom.) and absorption column
densities (10$^{20}\le $\NH\ $\le10^{22}$ \cmsq, from right to left).
From this figure it can be seen that most of the well defined colors lie within 
the area of a typical LMXB spectrum of $\Gamma=1.5-2.0$, with no
intrinsic absorption (e.g. \citet{Irwin_03,Fabbiano_06a}). 

\subsection{Variable and Transient Sources}
A characteristic of compact accretion sources such as LMXBs is
variability, and, as a result of the monitoring nature of the
observing campaign, we have been able to search for this behaviour in
NGC 3379. Out of the 132
sources, 53, 40\% of the sources within NGC 3379, have been defined as
variable. A further 18 sources are transient candidates; sources that either appear or disappear, or are only detected for a limited amount
of contiguous time during the observations and have a flux ratio between the
`on-state' and `off-state' of greater than 10 for transient
candidates (11 objects), and 5 for possible transients (7 objects). 

One of the
specific aims of our monitoring campaign has been to identity these transient
candidate sources as it has been suggested that field LMXBs are expected
to be transients \citep{Piro_02,King_02} and low luminosity
ultracompact binaries in GCs are also expected to be transient in
nature \citep{Bildsten_04}. This sub-population is investigated
in the forthcoming paper Brassington et al. (2007b, in preparation).

\subsection{The Absence of Low Luminosity GC-LMXBs}
From the optical data, 10 GC-LMXB associations have been found in NGC 3379. Nine of
these lie within the joint {\em Hubble/Chandra} field, where 53 field
LMXBs have also been detected. From comparing the cumulative XLFs of
these two populations, shown in Figure \ref{fig:lf}, we have found
that there is a significant lack of GC-LMXBs at luminosities
\LX$<\sim4\times10^{37}$~\ergps, with a KS test excluding that the two
distributions may be derived from the same parent population at
99.82\% confidence \citep{Fabbiano_07}. This result indicates that
there is a dearth of GC-LMXBs at lower luminosities, excluding a
single formation mechanism for {\em all} LMXBs, resolving a long
standing controversy in LMXB formation.

\vspace{-0.3cm}

\section{Future Work}
The full results of the work presented here are discussed in detail in the catalog
paper \citet{Brassington_07a}, and the GC-LMXB paper
\citet{Fabbiano_07}. In addition to these, further highlights
from the X-ray binary population of NGC 3379 will also be presented in
Brassington (2007b); an investigation into the transient
population of NGC 3379, and Brassington (2007c); a study of the
radial number density of LMXBs.
Forthcoming papers will also present: the properties of the ULX, the
X-ray luminosity function and the diffuse emission of the galaxy, as well
as the properties of the nuclear source and the spectral variability
of the luminous X-ray binary population.
The results from this deep observation will then be compared to the
X-ray source catalog of the old, GC rich elliptical galaxies NGC 4278,
which has also recently been the subject of a deep \CHANDRA\ observation.

\acknowledgements 

\footnotesize{The work presented in this paper has been the result of a
collaboration with G. Fabbiano, D.-W. Kim, A. Zezas, L. Angelini, R. L. Davies, T. Fragos, J. Gallagher, V. Kalogera, A. R. King, A. Kundu, S. Pellegrini, G. Trinchieri \& S. Zepf.
 This work was supported by {\em Chandra} G0 grant G06-7079A
(PI:Fabbiano) and subcontract G06-7079B (PI:Kalogera). We acknowledge
partial support from NASA contract NAS8-39073(CXC). A. Zezas
acknowledges support from NASA LTSA grant NAG5-13056. S. Pellegrini
acknowledges partial financial support from the Italian Space Agency
ASI (Agenzia Spaziale Italiana) through grant ASI-INAF I/023/05/0.}

\vspace{-0.3cm}

\bibliographystyle{mn2e}
\bibliography{nicky}

\begin{thebibliography}{}

\bibitem[\protect\citeauthoryear{{Bildsten} \& {Deloye}}{{Bildsten} \&
  {Deloye}}{2004}]{Bildsten_04}
{Bildsten} L. \&~{Deloye} C.~J.,  2004, ApJL, 607, L119

\bibitem[\protect\citeauthoryear{{Brassington}, {Fabbiano}, {Kim}, {Zezas},
  {Zepf}, {Kundu}, {Angelini}, {Davies}, {Gallagher}, {Kalogera}, {Fragos},
  {King}, {Pellegrini} \& {Trinchieri}}{{Brassington}
  et~al.}{2007}]{Brassington_07a}
{Brassington} N.~J., et~al., 2007,
  astro-ph/0711.1289

\bibitem[\protect\citeauthoryear{{Fabbiano}}{{Fabbiano}}{2006}]{Fabbiano_06a}
{Fabbiano} G.,  2006, ARAA, 44, 323

\bibitem[\protect\citeauthoryear{{Fabbiano}, {Brassington}, {Zezas}, {Zepf},
  {Angelini}, {Davies}, {Gallagher}, {Kalogera}, {Kim}, {King}, {Kundu},
  {Pellegrini} \& {Trinchieri}}{{Fabbiano} et~al.}{2007}]{Fabbiano_07}
{Fabbiano} G., et~al., 2007, astro-ph/0710.5126

\bibitem[\protect\citeauthoryear{{Fabbiano}, {Kim}, {Fragos}, {Kalogera},
  {King}, {Angelini}, {Davies}, {Gallagher}, {Pellegrini}, {Trinchieri}, {Zepf}
  \& {Zezas}}{{Fabbiano} et~al.}{2006}]{Fabbiano_06b}
{Fabbiano} G., et~al., 2006, ApJ, 650, 879

\bibitem[\protect\citeauthoryear{{Giacconi}}{{Giacconi}}{1974}]{Giacconi_74}
{Giacconi} R.,  1974, In {\em X-ray Astronomy}, Giacconi, R. \& Gursky, H.
  eds., p. 155, Dordrecht: Reidel

\bibitem[\protect\citeauthoryear{Grindlay}{Grindlay}{1984}]{Grindlay_84}
Grindlay J.~E.,  1984, Adv. Space Res., 3, 19

\bibitem[\protect\citeauthoryear{{Irwin}, {Athey} \& {Bregman}}{{Irwin}
  et~al.}{2003}]{Irwin_03}
{Irwin} J.~A.,  {Athey} A.~E. \&~{Bregman} J.~N.,  2003, ApJ, 393, 134

\bibitem[\protect\citeauthoryear{{King}}{{King}}{2002}]{King_02}
{King} A.~R.,  2002, MNRAS, 335, L13

\bibitem[\protect\citeauthoryear{{Piro} \& {Bildsten}}{{Piro} \&
  {Bildsten}}{2002}]{Piro_02}
{Piro} A.~L. \&~{Bildsten} L.,  2002, ApJL, 571, L103

\bibitem[\protect\citeauthoryear{{Verbunt} F. \&~{van den
  Heuvel}}{{Verbunt}}{1995}]{Verbunt_95}
{Verbunt} F. \&~{van den Heuvel} E. P.~J.,  1995, in {\em X-ray Binaries},
  Lewin, W. H. G., van Paradijs, J., van den Heuvel, E. P. J., eds., p. 457,
  Cambridge, UK: Cambridge Univeristy Press

\end{thebibliography}

\end{document}